\documentclass[prd,aps,twocolumn,amsmath,superscriptaddress,preprintnumbers]{revtex4}
\usepackage{graphicx}
\usepackage{dcolumn}
\usepackage{bm}
\usepackage{epsfig}
\usepackage{amsmath}
\input{epsf}
\usepackage{psfrag}
\usepackage{xcolor}
\usepackage{grffile}
\usepackage{graphicx}
\usepackage{multirow}
\usepackage{bm}
\usepackage{bbm}
\usepackage{color}
\usepackage{slashed}
\usepackage{braket}
\usepackage[capitalize]{cleveref}
\usepackage{comment}

\begin{document}

\title{ Test of Bell Locality Violation in Flavor Entangled Neutral Meson Pair }
\author{  Kaiwen Chen$^{1}$\footnote{kwchen@nnu.edu.cn}, Zhi-Peng Xing$^{1}$\footnote{zpxing@nnu.edu.cn}, Ruilin Zhu$^{1,2,3}$\footnote{rlzhu@njnu.edu.cn}}
\affiliation{
	$^1$ Department of Physics and Institute of Theoretical Physics, Nanjing Normal University, Nanjing, Jiangsu 210023, China\\
	$^2$ CAS Key Laboratory of Theoretical Physics, Institute of Theoretical Physics,
	Chinese Academy of Sciences, Beijing 100190, China\\
	$^3$ Peng Huanwu Innovation Research Center, Institute of Theoretical Physics,
	Chinese Academy of Sciences, Beijing 100190, China}

\begin{abstract}
	The quasi-spin entanglement of neutral mesons to test the Bell nonlocality is systematically studied. In the case of CP violation, the concrete expression of the Bell inequality for entangled neutral meson pairs is derived. The violation of Bell inequality in the oscillations of  $B_d^0-\bar{B}_d^0$, $B_s^0-\bar{B}_s^0$, $D^0-\bar{D}^0$ and $K^0-\bar{K}^0$ meson pair systems are found at certain evolution time including
the latest experimental data.  Our study is helpful for studying CP-violating involved flavor entanglement and testing Bell inequality violation in the current and future high energy experiment facilities.
\end{abstract}

\maketitle

\section{Introduction}

Quantum entanglement is a unique phenomenon in quantum mechanics (QM), which does not exist in the classical physics. Entanglement in quantum mechanics causes these correlative systems to have instantaneous effects on each other regardless of distance. In 1935, Einstein, Podolsky and Rosen (EPR)~\cite{Einstein:1935rr} proposed a hypothetical experiment to demonstrate that quantum mechanics is incomplete and introduce hidden variable local theory to solve this problem. However, this idea remained untestable until 1965, when Bell proposed the Bell inequality~\cite{Bell:1964kc}, which local realism (LR) would need to satisfy and quantum mechanics would violate under certain circumstances.

The violation of the Bell inequality was initially detected by low-energy photon pairs with spin entanglement~\cite{Aspect:1982fx,Weihs:1998gy}. The two photons are prepared into a singlet state, and the polarizations of the pair of photons are then measured from different directions~\cite{Horodecki:2009zz}, which proves that quantum mechanics breaks the Bell inequality. Many experiments were further performed  using superconducting circuits~\cite{Storz:2023jjx}, and entangled
atoms~\cite{Rosenfeld:2017rka}.

Recently, the measurement of quantum entanglement and the test of Bell inequality violation have reattracted a lot of attention in the field of high energy particle physics. In high energy physics, quantum entanglement and the violation of Bell inequality are tested mainly through spin-correlative particle pairs. For a two-qubit system, Bell nonlocality is measured mainly by Clauser-Horne-Shimony-Holt (CHSH) inequality~\cite{Clauser:1969ny}. There have been a lot of works and results in high energy physics for the testing of such systems, e.g., $e^+e^- \to Y\bar{Y}$ at BESIII~\cite{Wu:2024asu,Li:2006fy}, $e^+e^- \to \tau^+\tau^-$ at BESII~\cite{Ehataht:2023zzt}, through top-quark, tau-lepton and photon pairs~\cite{Fabbrichesi:2022ovb}, and the spinning gluons system~\cite{Guo:2024jch}.
 For a two-qutrit system, the most commonly used inequality for Bell nonlocality detection is the Collins-Gisin-Linden-Massar-Popescu (CGLMP) inequality~\cite{Collins:2002sun}, which can be generalized to systems of arbitrary dimensions. Great progresses have also been made in  quantum entanglement and Bell non-local measurement of such systems, e.g.,  the $B$ meson decays into vector mesons~\cite{Fabbrichesi:2023idl,Li:2009rta}, the $B_c$ meson decays into vector mesons~\cite{Chen:2024syv,Geng:2023ffc}, and the $H \to Z Z$ decays with anomalous coupling~\cite{Bernal:2024xhm,Aguilar-Saavedra:2022wam}.

In this work, we are mainly interested in the measurement of Bell nonlocality of entangled neutral meson pairs. We can view the flavor bases of these mixing particles as quasi-spins of the two-body mixing sytem. The flavor eigenstate is generally not the eigenstate of the neutral meson energy, so the polarization direction of the particle in the quasi-spin space will change with time. So we can choose different time points to measure the entangled neutral meson pair just as we can choose different directions to measure the polarization of a particle with spin. This method to test the Bell non-locality has been demonstrated by many previous literatures~\cite{Benatti:1997xt,Foadi:1999sg,Foadi:2000zz,Samal:2002mh,Ancochea:1998nx,Bertlmann:2001ea,Bramon:2002yg,Go:2003tx,Bramon:2005mg,Bertlmann:2006hs,Takubo:2021sdk}.
Therein  the violation of the CHSH inequality is observed in  $B^0-\bar{B}^0$ system using semileptonic $B^0/\bar{B}^0$ decays data at Belle experiment~\cite{Go:2003tx}. However, the CP violation effects in quantum entanglement are not included in these works, which motivates
us to systematically study the quasi-spin entanglement of neutral meson pairs in the most general case. Though the CP violation effects in neutral meson pair mixing are tiny, the precision particle experiments shall help us to capture the nature of quantum flavor entanglement  and the nature of CP violation. In addition, the multi-dimensional time evolution and the maximum breaking of Bell inequality is also given.

The paper is arranged as follows. In Sec.~II, we derive the evolution of the flavor wave function with time in the most general case of neutral entangled mesons, and consider the effect of CP violation on the evolution of the wave function. In Sec.~III, we describe how to measure the Bell nonlocality of entangled neutral meson pairs and give the time-dependent CHSH inequality firstly. Then we calculate the specific expression of the CHSH inequality after the probability normalization at a specific point in time. Next we give out the time evolution of CHSH inequality value  for different kinds of neutral meson pair system $B_d^0-\bar{B}_d^0$, $B_s^0-\bar{B}_s^0$, $D^0-\bar{D}^0$ and $K^0-\bar{K}^0$. A brief summary is given in the end.

\section{Evolution of neutral mesons over time}
Considering the instantaneous production of a neutral meson pair in which two neutral mesons form a matter-antimatter pair, the neutral meson pair usually make up a flavor singlet through electromagnetic and strong interactions.  Take the production of neutral $B$ meson pair for example, $\Upsilon(4S)$ are largely produced at Belle-II experiments by the electro-positron annihilation and then decay into a $B_d^0\bar{B}_d^0$ pair. In the rest frame of $\Upsilon(4S)$, a $B_d^0\bar{B}_d^0$ pair will move in opposite directions with the same magnitude of momentum. Since the $J^{PC}$ of the $\Upsilon(4S)$ equals $1^{--}$ and the  parity is conserved in the decay process, the flavor space wave function of the $B_d^0\bar{B}_d^0$ pair should be antisymmetric~\cite{Banerjee:2014vga}. The same situation can be extended to the $D$ and $K$ meson pair systems. So the flavor wave function of this kind of neutral meson pair at the initial time $t=0$ can be written as
\begin{align}
	\ket{\psi(0)}=\frac{1}{\sqrt{2}}(\ket{M\bar{M}}-\ket{\bar{M}M}),
\end{align}
where $M$ represents one kind of $B_d^0$, $B_s^0$, $K^0$, and $D^0$ particles while $\bar{M}$ represents the anti-particle. $\ket{M}$ and $\ket{\bar{M}}$ are flavor eigenstates of neutral mesons. $\ket{M\bar{M}}$ denotes the case that one particle $M$ flies to the right side and the other particle $\bar{M}$ flies to the left side. It is clear that the initial wave function is maximally flavor entangled state.

However, flavor eigenstates are generally not energy eigenstates of neutral mesons. In order to study the evolution of the wave function over time, we need to express the wave function as a superposition of energy eigenstates. In general, the energy eigenstate of a neutral meson can be written as
\begin{align}
	&\ket{M_1}=p\ket{M}+q\ket{\bar{M}}, \\
	&\ket{M_2}=p\ket{M}-q\ket{\bar{M}}.
\end{align}
For example, it is known as $\ket{M_1}=\ket{B_L}$ for light $B$ meson and $\ket{M_2}=\ket{B_H}$ for heavy $B$ meson. In the above formulation, in order to normalize the wave function it requires $|p|^2+|q|^2=1$. We have  $p=q=\frac{1}{\sqrt{2}}$ in the case of CP conservation while $p\neq q$ in the case of CP violation.

 The evolution of energy eigenstates can be expressed as~\cite{Bramon:2005mg}
\begin{align}
	&\ket{M_1(t)}=e^{-im_1t} e^{-\frac{1}{2}\Gamma_1t} \ket{M_1}, \\
	&\ket{M_2(t)}=e^{-im_2t} e^{-\frac{1}{2}\Gamma_2t} \ket{M_2},
\end{align}
where we have considered the decay properties of neutral mesons. $m_i$ and $\Gamma_i$ represents the mass and the decay width of $M_i$($i=1,2$) meson.

If $t_1$ and $t_2$ are used to represent the evolution time of the left and right side particles respectively, the wave function of the neutral meson pair at a particular moment can be written as
\begin{align}
	\ket{\psi(t_1,t_2)}=
	&\frac{1}{2\sqrt{2}pq}e^{-i(m_2t_1+m_1t_2)-\frac{1}{2}(\Gamma_2t_1+\Gamma_1t_2)}\big(\ket{M_2M_1} \nonumber \\
	&-e^{i\Delta m(t_1-t_2)+\frac{1}{2}\Delta\Gamma(t_1-t_2)}\ket{M_1M_2}\big) ,
\end{align}
where $\Delta m= m_2-m_1$ and $\Delta \Gamma=\Gamma_2-\Gamma_1$. It is well known that the measurement of flavor eigenstates is a good way that can be measured well experimentally through their decay products. So changing the above formula to flavor eigenstates, one can find that
\begin{align}
	\ket{\psi(t_1,t_2)}=&\frac{B}{2\sqrt{2}pq}
	\big[p^2(1-A)\ket{MM}+pq(1+A)\ket{M\bar{M}} \nonumber \\
	&-pq(1+A)\ket{\bar{M}M}-q^2(1-A)\ket{\bar{M}\bar{M}}.
\end{align}
For simplicity, we have defined that
\begin{align}
	&A=e^{i\Delta m(t_1-t_2)+\frac{1}{2}\Delta\Gamma(t_1-t_2)} , \\
	&B=e^{-i(m_2t_1+m_1t_2)-\frac{1}{2}(\Gamma_2t_1+\Gamma_1t_2)}.
\end{align}
The wave function of the neutral meson  pair at a given time can then be obtained by simply inserting the parameters of the specific neutral meson into Eq.~(7).

\section{Bell inequality violation in neutral mesons}

 The most commonly used extension of Bell inequality for measuring a pair of particles with spin $\frac{1}{2}$, the CHSH inequality~\cite{Clauser:1969ny}, can be expressed as
\begin{align}
	S=|E(\vec{a},\vec{b})+E(\vec{a}^{\prime},\vec{b})+E(\vec{a},\vec{b}^{\prime})-E(\vec{a}^{\prime},\vec{b}^{\prime})| \le 2.
\end{align}
In the above formula, $\vec{a}$ and $\vec{a}^{\prime}$ mean to choose two different directions to measure the polarization of one of the particles similar to $\vec{b}$ and $\vec{b}^{\prime}$ for the other particle. The detection of Bell inequality for neutral mesons is somewhat different from the detection for polarized particles.

For neutral mesons, their flavor bases can be regarded as their quasi-spin. The direction of the quasi-spin of neutral mesons changes with time. One can choose four different time points to measure the neutral meson pair, similarly to choose four different directions to measure the spinning particle pair. The CHSH inequality is expressed as
\begin{align}
	S=|E(t_1,t_2)+E(t_1^{\prime},t_2)+E(t_1,t_2^{\prime})-E(t_1^{\prime},t_2^{\prime})| \leq 2,
\end{align}
where the correlation function $E(t_1,t_2)$ is defined as
\begin{align}
	E(t_1,t_2)=
	&P(t_1,M;t_2,M)+P(t_1,\bar{M};t_2,\bar{M}) \nonumber \\
	&-P(t_1,M;t_2,\bar{M})-P(t_1,\bar{M};t_2,M).
\end{align}
The joint probability density $P(t_1,X;t_2,X')$ is the probability that $X$ produced in the left side is measured at $t_1$ while $X'$ produced in the right side is measured at $t_2$.

Using Eq.~(7) and defining $C=\frac{|p|^2}{|q|^2}$, it is easy to calculate the joint probability density at time $(t_1,t_2)$
\begin{align}
	&P(t_1,M;t_2,M)=\frac{C}{8}[e^{-(\Gamma_2t_1+\Gamma_1t_2)}+e^{-(\Gamma_1t_1+\Gamma_2t_2)}] \nonumber \\
	&~~~ ~~~ ~~~ ~~~ ~~~ ~~~ ~~~ ~~~[1-\frac{cos\Delta m(t_1-t_2)}{cosh\frac{\Delta\Gamma(t_1-t_2)}{2}}], \\
	&P(t_1,\bar{M};t_2,\bar{M})=\frac{1}{8C}[e^{-(\Gamma_2t_1+\Gamma_1t_2)}+e^{-(\Gamma_1t_1+\Gamma_2t_2)}] \nonumber \\
	&~~~ ~~~ ~~~ ~~~ ~~~ ~~~ ~~~ ~~~[1-\frac{cos\Delta m(t_1-t_2)}{cosh\frac{\Delta\Gamma(t_1-t_2)}{2}}], \\
	&P(t_1,M;t_2,\bar{M})=\frac{1}{8}[e^{-(\Gamma_2t_1+\Gamma_1t_2)}+e^{-(\Gamma_1t_1+\Gamma_2t_2)}] \nonumber \\
	&~~~ ~~~ ~~~ ~~~ ~~~ ~~~ ~~~ ~~~[1+\frac{cos\Delta m(t_1-t_2)}{cosh\frac{\Delta\Gamma(t_1-t_2)}{2}}], \\
	&P(t_1,\bar{M};t_2,M)=\frac{1}{8}[e^{-(\Gamma_2t_1+\Gamma_1t_2)}+e^{-(\Gamma_1t_1+\Gamma_2t_2)}] \nonumber \\
	&~~~ ~~~ ~~~ ~~~ ~~~ ~~~ ~~~ ~~~[1+\frac{cos\Delta m(t_1-t_2)}{cosh\frac{\Delta\Gamma(t_1-t_2)}{2}}].
\end{align}
Then one need to normalize the probability at time $(t_1,t_2)$, otherwise one can not observe the violation of Bell inequality due to the particle decays. One can normalize the joint probability density by the formula~\cite{Takubo:2021sdk}
\begin{align}
	P_{t_1,t_2}(X,X^{\prime})=\frac{P(t_1,X;t_2,X^{\prime})}{\sum_{X,X^{\prime}}P(t_1,X;t_2,X^{\prime})},
\end{align}
where $X$ and $X^{\prime}$ denote $M$ or $\bar{M}$. After normalization, the expression for the joint probability density can be written as
\begin{align}
	&P_{t_1,t_2}(M,M)=\frac{C^2(1-F)}{(C+1)^2-(C-1)^2F}, \\
	&P_{t_1,t_2}(\bar{M},\bar{M})=\frac{(1-F)}{(C+1)^2-(C-1)^2F}, \\
	&P_{t_1,t_2}(M,\bar{M})=\frac{C(1+F)}{(C+1)^2-(C-1)^2F}, \\
	&P_{t_1,t_2}(\bar{M},M)=\frac{C(1+F)}{(C+1)^2-(C-1)^2F},
\end{align}
where
\begin{align}
	F=\frac{\cos\Delta m(t_1-t_2)}{\cosh\frac{\Delta\Gamma(t_1-t_2)}{2}}.
\end{align}

In combination with formula in Eq.~(12), the expression of the correlation function including the CP violation effects can be obtained as
\begin{align}
	E(t_1,t_2)=\frac{(C-1)^2-(C+1)^2F}{(C+1)^2-(C-1)^2F}.
\end{align}
It is obvious that there will be four completely independent time variables in the expression of CHSH inequality $S$ at the end, which makes the experimental measurement more complex. So in the following, we will fix the value of $t_1$, $t_2$ and let $S$ evolve as a function of other two time potints $t_1^{\prime}$ and $t_2^{\prime}$. In addition, it is best to measure the point in time as quickly as possible due to the existence of decay of the neutral mesons. So we will focus on the earliest $S$ maximum value. Of course, due to relativistic time dilatation effect, the neutral mesons can survive for a while to be detected if they are produced with a higher energy.

\begin{table}[h!]
	\caption{Latest data for neutral mesons~\cite{ParticleDataGroup:2022pth,Krzemien:2024eep}.}
	\begin{tabular}{|c|c|c|c|c|}
		\hline
		Particles & Lifetime(ps) & \textbf{$\Delta m$}(ps$^{-1}$) & \textbf{$\Delta \Gamma$}(ps$^{-1}$) & \textbf{$C$}  \\
		\hline
		$B_d^0$,$\bar{B}_d^0$ & $1.517\pm0.004$ &0.5069  & $\sim$ 0  & 0.998 \\
		\hline
		$B_s^0$,$\bar{B}_s^0$  &$1.520\pm0.005$ & 17.765 & 0.083  & 0.9994 \\
		\hline
		$D^0$,$\bar{D}^0$  &$0.410\pm0.001$ & 0.00993  & 0.03148  & 1.0121 \\
\hline
		$K_S$,  &
                  $89.54\pm0.04$,
& \multirow{2}{*}{ 0.005292 } & \multirow{2}{*}{0.01115  }& \multirow{2}{*}{1.0064} \\	
		$K_L$  &
                  $51160\pm210$
&  & &  \\	
		\hline
	\end{tabular}
	\label{table}
\end{table}

\begin{figure}[thp]
	\includegraphics[width=0.45\textwidth]{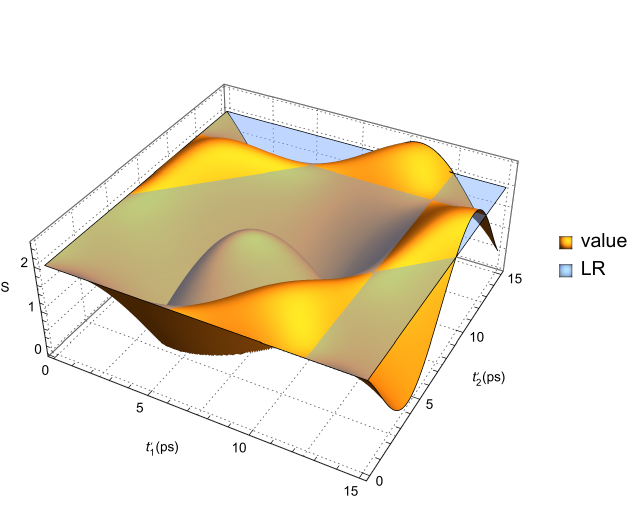}
	\caption{The CHSH inequality $S$ for $B^0_d-\bar{B}^0_d$ system as a function of $t_1^{\prime}$ and $t_2^{\prime}$ with $t_1=t_2=0$ ps.}
	\label{fig:Bd2}
\end{figure}

\begin{figure}[thp]
	\includegraphics[width=0.45\textwidth]{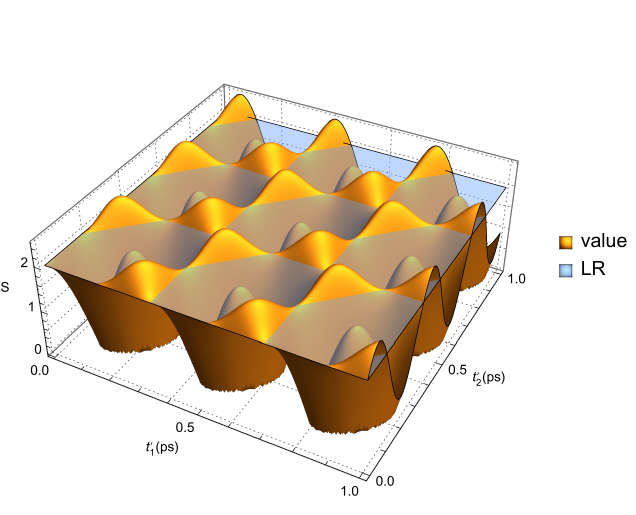}
	\caption{The CHSH inequality $S$ for $B^0_s-\bar{B}^0_s$ system as a function of $t_1^{\prime}$ and $t_2^{\prime}$ with $t_1=t_2=0$ ps.}
	\label{fig:Bs2}
\end{figure}

\begin{figure}[thp]
	\includegraphics[width=0.45\textwidth]{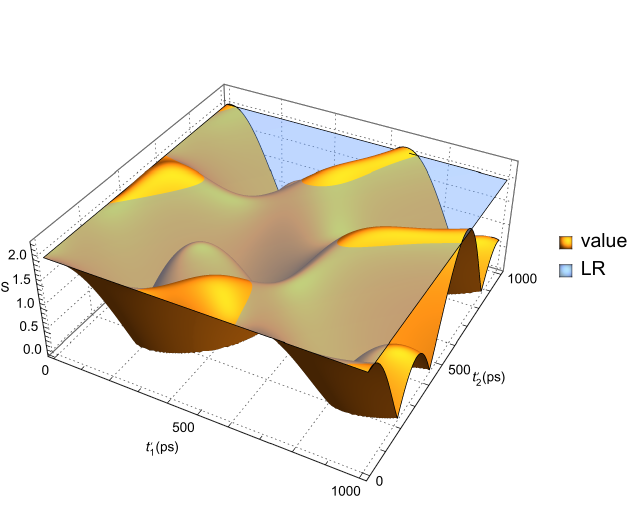}
	\caption{The CHSH inequality $S$ for $D^0-\bar{D}^0$ system as a function of $t_1^{\prime}$ and $t_2^{\prime}$ with $t_1=t_2=0$ ps.}
	\label{fig:D2}
\end{figure}

\begin{figure}[thp]
	\includegraphics[width=0.45\textwidth]{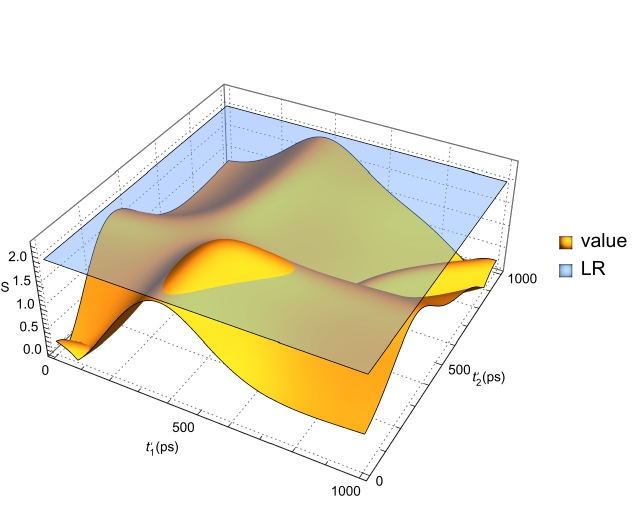}
	\caption{The CHSH inequality $S$ for $K^0-\bar{K}^0$ system as a function of $t_1^{\prime}$ and $t_2^{\prime}$ with $t_1=300$ ps and $t_2=372$ ps.}
	\label{fig:K2}
\end{figure}

 In the calculation, the relevant parameter inputs for the neutral mesons are listed in Table.~\ref{table}. Fixing two time points and varying two other time points, the time evolution of CHSH inequality S for different neutral meson pair system over time is shown in~\cref{fig:Bd2,fig:Bs2,fig:D2,fig:K2} respectively. The yellow plane represents the value of $S$ over time. The blue translucent plane represents the maximum value predicted by local realism (LR). There are two kinds of time dependence for CHSH inequality S. The first is a periodic change in the value of $S$ caused by factor $cos\Delta m(t_1-t_2)$. The second is the depression of the maximum value of the adjacent period $S$ caused by factor $cosh\frac{\Delta\Gamma(t_1-t_2)}{2}$.

For simplicity, we can fix the initial time $t_1=t_2=0$ ps, naming we first observe the neutral meson pair produced instantaneously.
In the case of  neutral $B^0_d-\bar{B}^0_d$ system, the value of $S$ reaches its maximum value 2.5 for the first time with $t_1^{\prime}=10.3$ ps and $t_2^{\prime}=2.1$ ps, or swap the two time values because of the symmetry of the two times. In addition, the maximum value of $S$ in each oscillation period does not decay with time because of $\Delta \Gamma_{B^0_d}=0$. For $B^0_s-\bar{B}^0_s$ system,  the value of $S$ reaches its maximum 2.5 for the first time at $t_1^{\prime}=0.29$ ps, $t_2^{\prime}=0.06$ ps. One can see that the oscillation period of the $S$ value of $B^0_s-\bar{B}^0_s$ system is very short because the value of $\Delta m_{B_s^0}$ is very large. At the same time, the depression of the maximum value of $S$ between adjacent periods can hardly been reflected due to $\Delta m_{B_s^0} \gg \Delta \Gamma_{B_s^0}$ unless the evolution of $S$ is plotted on a larger time scale. For $D^0-\bar{D}^0$ system, one can find that the first time that $S$ reaches its maximum 2.3 is at $t_1^{\prime}=496$ ps and $t_2^{\prime}=80.3$ ps. For $K^0-\bar{K}^0$ sytem, the situation will be quite different. Because the value of $\Delta m_{K}$ is very small, the oscillation period of $K^0$ is very long. Meanwhile, $\Delta \Gamma_{K^0} \approx 2\Delta m_{K^0}$ causes a significant decrease in the maximum value of $S$ between adjacent periods. In this case, only a very small number of peaks are sufficient to violate CHSH inequality. After searching, we fix the time $t_1=300$ ps and $t_2=372$ ps. The only peak of $S$ that violates Bell inequality is 2.34 at the time $t_1^{\prime}=444.9$ ps, $t_2^{\prime}=227.1$ ps.

If we vary all the four measurement point for quasi-spin, the value of $S$ is 2.18 with $t_1=1$ ps, $t_2=0.5$ ps, $t'_1=0$ ps and $t'_2=1.5$ ps,  while the maximum value of $S$ becomes 2.83 with $t_1=1.55$ ps, $t_2=3.10$ ps, $t'_1=0$ ps and $t'_2=4.65$ ps for $B^0_d-\bar{B}^0_d$ system.  For  $B^0_s-\bar{B}^0_s$ system, the maximum value of $S$ becomes 2.83 with $t_1=0.50$ ps, $t_2=0.54$ ps, $t'_1=0.59$ ps and $t'_2=0.45$ ps. For  $D^0-\bar{D}^0$ system, the maximum value of $S$ becomes 2.77 with $t_1=80.6$ ps, $t_2=156.4$ ps, $t'_1=232.2$ ps and $t'_2=4.71$ ps.  For  $K^0-\bar{K}^0$ system, the value of $S$ is 2.13 with $t_1=29.7$ ps, $t_2=59.3$ ps, $t'_1=89$ ps and $t'_2=0$ ps,  while the maximum value of $S$ becomes 2.35 with $t_1=205$ ps, $t_2=278$ ps, $t'_1=351$ ps and $t'_2=133$ ps. Note that the choices of time measurement points are not unique when the CHSH inequality is max broken. In general, the violation of the Bell inequality generally exists in the above processes as long as the appropriate time observation point is chosen. By choosing these appropriate time observation point, the experiment collaboration such as Belle-II, BESIII and future Super tau-charm factories(STCF) and  Circular Electron Positron Collider(CEPC) can test the Bell inequality and explore what extent the Bell locality is broken~\cite{Achasov:2023gey,CEPCStudyGroup:2023quu}. Among these neutral meson pair, the test of Bell locality violation in $D^0-\bar{D}^0$ system is relatively difficult compared with  other neutral meson entanglement systems in experiments due to its short lifetime and narrow mass splitting.

\section{conclusion}
In this paper, the time evolution of the entangled wave function of neutral meson pair  is derived theoretically where minor CP violation parameters are also taken into account. In order to investigate the violation of Bell inequality during the oscillation of neutral mesons, the concrete expression of CHSH inequality including CP violation effect is calculated. By studying the oscillations of mesons $B_d^0$, $B_s^0$, $D^0$ and $K^0$, we find that there are violations of Bell inequality in these processes as long as the appropriate time node is selected to measure. The maximum breaking for CHSH inequality is obtained. It shows that these oscillations of neutral mesons accord with the prediction of quantum mechanics (QM) and cannot be explained by local realism (LR). This makes a contribution to test the violation of Bell inequality in high-energy physics. It also helps us better understand flavor entanglement and Bell non-locality in particle physics.

\section*{Acknowledgments}
The work is  supported by NSFC under grant Nos. 12322503, 12075124, 12375088 and 12335003 and by Natural Science Foundation of Jiangsu under Grant No. BK20211267.

\end{document}